\documentclass[a4paper,11pt]{article}
\pdfoutput=1 

\usepackage{jheppub} 
\usepackage[T1]{fontenc} 

\usepackage{feynmf}

\def\eps{\varepsilon}
\def\nn{\nonumber}
\def\[#1\]{\begin{align}#1\end{align}}

\def\uv{{\rm UV}}

\def\td#1{\tilde{\delta}\left(#1\right)}
\def\ii{\imath 0}
\def\pb{\mathbf{p}}
\def\lb{\boldsymbol{\ell}}

\def\eq#1{eq.~(\ref{#1})}
\def\reff#1{ref.~\cite{#1}}
\def\ln#1{\log \left(#1\right)}

\title{Asymptotic expansions through the loop-tree duality}

 \author{Judith Plenter}
 \author{and Germ\'an Rodrigo}
 \affiliation{Instituto de F\'{\i}sica Corpuscular, Universitat de Val\`{e}ncia -- 
Consejo Superior de Investigaciones Cient\'{\i}ficas, Parc Cient\'{\i}fic, E-46980 Paterna, Valencia, Spain}

\emailAdd{plenter@ific.uv.es}
\emailAdd{german.rodrigo@csic.es}

\abstract{
First results towards a general method for asymptotic expansions of Feynman amplitudes in the loop-tree duality (LTD) formalism are presented. The asymptotic expansion takes place at integrand-level in the Euclidean space of the loop three-momentum, where the hierarchies among internal and external scales are well-defined. Additionally, the UV behaviour of the individual contributions to the asymptotic expansion emerges only in the first terms of the expansion and is renormalized locally in four space-time dimensions. These two properties represent an advantage over the method of Expansion by Regions (EbR). We explore different approaches  in different kinematical limits, and derive general guidelines with several benchmark examples.}


\begin{document} 
\maketitle
\flushbottom

\section{Introduction}

Since the Higgs boson has been discovered succesfully at CERN's Large Hadron Collider (LHC) almost one decade ago, clear evidence of further particles not described by the Standard Model (SM) has not been reported, although observations such as the muon's anomalous magnetic moment or the $B$ anomalies hint at discrepancies between SM predictions and measurements. 
With decreasing experimental uncertainties it is clear that theoretical predictions at increasing precision are at the forefront of current research in order to confirm or reject deviations from the SM. 
Consequently, higher order contributions in perturbative Quantum Field Theory (pQFT) are crucial. This endeavour quickly reaches its limits in the common approach of Dimensional Regularization (DREG). Within this technique the divergent expressions that appear in loop calculations of Feynman diagrams are regularized by working in $d=4-2\eps$ space-time dimensions, thus leaving the problematic integrals formally well-defined. The limit $\eps\to 0$ is taken only after both infrared (IR) and ultraviolet (UV) singularities have been canceled in IR safe observables and/or renormalized through appropriate redefinition of the constants appearing in the theory. Within DREG, the difficulty posed by the integral(s) in a Feynman amplitude scales with the number of loops, external legs and mass scales. 

The interest in asymptotic expansions within pQFT arises from their potential to facilitate analytic results in specific kinematic configurations, particularly when full analytic calculations in DREG are not possible. While work on the solution for further master integrals is ongoing, an expanded result can still be of great interest since it showcases the relevant behaviour of the amplitude in the needed kinematic limit. There are many observables where an analytic result is not necessary for every set of kinematics and where specific limits are the window to test potential discrepancies between experiments and SM predictions, thus identifying new physics contributions. Furthermore, in the context of the local cancellation of IR singularities expanded integrands could be very convenient to reduce computation time.

As an example for a new-physics scenario highly-boosted Higgs boson production may be mentioned: while the regime of small transverse momentum has been calculated with a point-like interaction encoding the top-quark loop \cite{Chen:2014gva,Boughezal:2015dra}, first attempts at the full calculation necessary for obtaining the large transverse momentum distribution have only been published recently and rely on either numerical integration \cite{Jones:2018hbb} or expansions in the Integration by Parts identities \cite{Lindert:2018iug}. It is exactly this part of the amplitude which is needed in order to rule out an additional point-like effective Higgs-gluon-gluon coupling. 

The interest in asymptotic expansions becomes also clear noting that there are already well-developed methods for simplifying the integrands of Feynman amplitudes. Widely known among them is Expansion by Regions \cite{Beneke:1997zp, Smirnov:2002pj,Pak:2010pt,Jantzen:2011nz,Jantzen:2012mw,Mishima:2018olh}. While this technique has been shown to provide correct results a general proof is still pending \cite{Semenova:2018cwy}. Additionally, the degree of UV divergence rises with every term in the expansion which can be considered inconvenient.

In recent years an alternative regularization method based on the loop-tree duality (LTD) has been developed and applied both at one loop and beyond \cite{Catani:2008xa, Bierenbaum:2010cy, Bierenbaum:2012th, Buchta:2014dfa, Buchta:2015xda, Buchta:2015wna, Hernandez-Pinto:2015ysa, Sborlini:2016gbr, Sborlini:2016hat, Driencourt-Mangin:2017gop, Tomboulis:2017rvd, Driencourt-Mangin:2019aix, Driencourt-Mangin:2019sfl, Runkel:2019yrs,Baumeister:2019rmh,Capatti:2019ypt,Capatti:2019edf,Aguilera-Verdugo:2019kbz,Verdugo:2020kzh}. There are other alternative methods to DREG which are summarized in Ref.~\cite{Gnendiger:2017pys}. The basis of LTD is using the Cauchy residue theorem to integrate one component of the loop momentum. Loop amplitudes can thus be expressed as a sum of residues which can be reformulated as so-called dual amplitudes. These consist of sums of tree-level like objects to be integrated in what essentially is a phase-space integral.

Since as a result of LTD one obtains a function to be integrated over a Euclidean three-momentum it is possible to cancel IR singularities locally. This feature allowed the development of the Four-dimensional Unsubtraction method (FDU) \cite{Hernandez-Pinto:2015ysa, Sborlini:2016gbr, Sborlini:2016hat}. Further, it leads to an additional characteristic: in comparison to the original amplitude as a function of Minkowski four-momenta the size of scalar products appearing in the dual integrand can be directly compared to external scales. This allows the development of a well-defined formalism of asymptotic expansions of the integrand. Some successful results of expansions in the context of LTD have already been achieved in the process $H\to\gamma\gamma$ at one loop \cite{Driencourt-Mangin:2017gop}. The aim of this paper is the introduction of a general formalism. First steps have been reported on in \cite{Plenter:2019jyj} recently.

In this work we present the starting point for the development of a general method for asympotic expansions in the context of LTD. General guidelines for the expansion of the dual propagator are layed out in section \ref{sec:dualprop}. Those rules are then applied to the bubble diagram in section \ref{sec:bubble} as well as the scalar three-point function in section \ref{sec:scalar3pt}, in both cases for a variety of limits. We aim towards obtaining an expansion that is well-defined also at integrand-level and simplifies integrands sufficiently to obtain loop analytic results at higher orders and multiple scales. One of the long-term goals of our work will be to obtain an independent calculation at two-loops of Higgs boson production with large transverse momentum by expanding the integrand in this regime.

\section{Loop-tree duality and asymptotic expansions of dual propagators} 
\label{sec:dualprop}

A general one-loop scattering amplitude with $N$ external legs in the Feynman representation is given by
\[
\def\uv{{\rm UV}}
	\mathcal{A}^{(1)}_N &= \int_\ell \mathcal{N}\left( \ell, \{p_k\}_N \right) \left( \prod_{i=1}^N G_F(q_i) \right)~, 
	\label{eq:twopointfeynman}
\] 
where the integral measure in $d=4-2\eps$ space-time dimensions is $\int_\ell = -\imath \,\mu^{4-d} \int \mathrm{d}^d \ell / (2\pi)^d$,
$\mathcal{N}\left( \ell, \{p_k\}_N \right)$ is a function of the loop momentum $\ell$ and the $N$ external momenta $\{p_k\}_N$.
$G_F(q_i) = (q_i^2-m_i^2 + \ii)^{-1}$ is the Feynman propagator carrying momentum $q_i = \ell + k_i$, where $k_i$ a linear combination of the external momenta. Applying the loop-tree duality theorem this amplitude is rewritten as 
\[
	\mathcal{A}^{(1)}_N &= - \int_\ell \mathcal{N} \left( \ell, \{p_k\}_N \right) \sum_{i=1}^N \td{q_i} \left( \prod_{j\neq i} G_D \left( q_i; q_j \right) \right)~, 
	\label{eq:LTDformula}  
\]
where $G_D (q_i;q_j) = ( q_j^2- m_j^2 - \ii \, \eta\cdot k_{ji})^{-1}$, with $k_{ji} = q_j-q_i$, are the so-called dual propagators and $\eta$ is an arbitrary future-like vector. The dual propagators differ from the Feynman propagators only in their infinitesimal imaginary prescription, whose sign in the dual propagator depends on the external momenta. 
A different internal loop momentum is set on-shell in each of the terms in \eq{eq:LTDformula}, which are conventionally called dual amplitudes, through the modified delta functional $\td{q_i} = 2\pi i ~ \theta(q_{i,0}) \delta (q_i^2 - m_i^2 )$, in short, or $\td{q_i; m_i} \equiv \td{q_i}$ whenever it is necessary to make reference to different internal masses. Due to the on-shell conditions, the dimensions
of the integration domain is reduced by one unit. The choice $\eta=(1,\mathbf{0})$ is the most convenient because it is equivalent to integrating out the energy component of the loop momentum,  thus reducing the integration measure to the Euclidean space of the loop three-momentum.  

The behaviour of scattering amplitudes is ruled by their analytic properties. Aiming for asymptotic expansions at integrand-level, we must therefore consider in detail the analysis of propagators which are the objects that give rise to singularities. While the numerator plays a role in determining whether the amplitude has a UV divergence this is insignificant for asymptotic expansions since within LTD the singular UV behaviour is neutralized through local renormalization before integration. An example of this will be shown in the following section. 

The dual propagators can manifest non-causal or unphysical singularities on top of the physical divergences related to causal threshold and IR singularities. These unphysical divergences appear only when the various terms in the sum are considered separately. Identifying the conditions under which both causal and unphysical singularities appear as well as their position in the integration space is necessary groundwork for asymptotically expanding an amplitude. The examination of said singularities can be achieved efficiently by reparametrizing the dual propagators as shown in \reff{Buchta:2014dfa,Aguilera-Verdugo:2019kbz} 
\[
 	\frac{\td{q_i}}{\pi \imath} \, G_D (q_i;q_j) = 
	\frac{\delta \left( q_{i,0}-q_{i,0}^{(+)} \right)}{q_{i,0}^{(+)} \lambda_{ij}^{+-} \lambda_{ij}^{++}}, \qquad \lambda_{ij}^{\pm \pm} = \pm q_{i,0}^{(+)} \pm q_{j,0}^{(+)} + k_{ji,0} ~,
\]
where $q_{i,0}^{(+)} = \sqrt{{\bf q}_i^2+m_i^2}$ are the on-shell energies.
In this notation, a causal unitarity threshold appears for $\lambda_{ij}^{++}\to 0$ while an unphysical singularity appears for $\lambda_{ij}^{+-}\to 0$. The latter case always appears entangled between two dual amplitudes which leads to the cancellation of these unphysical singularities due to the always opposite sign of the infinitesimal imaginary prescription in the dual propagators. It is straightforward to derive the kinematic conditions for either of these limits to occur and examples are provided in \reff{Buchta:2014dfa,Aguilera-Verdugo:2019kbz}. In some special kinematic configurations, the unphysical singularities may even be avoided altogether by redefining the loop momentum flow through $\ell\to -\ell$,
see e.g. \reff{Aguilera-Verdugo:2019kbz}. Remarkably, we have recently presented dual representations of selected multiloop topologies that are explicitly free of 
unphysical singularities, and we conjectured that this property holds to other loop topologies at all orders \cite{Verdugo:2020kzh}. The advantages that this other representation introduces will be explored further in future publications. 

Having identified the propagators of the amplitude that lead to singularities, we can now reparametrize the dual propagators in the following form that 
is more suitable for asymptotic expansions
\[
	\td{q_i} \, G_D \left( q_i ; q_j\right) &= \frac{\td{q_i}}{2q_i\cdot k_{ji} + \Gamma_{ij}+\Delta_{ij} - \ii \eta\cdot k_{ji}}~,
	\label{eq:dualpropagator}
\]
where $\Gamma_{ij}+\Delta_{ij} = k_{ji}^2 + m_i^2 - m_j^2$. If $\Gamma_{ij}+\Delta_{ij}$ vanishes the dual propagator is not expanded. Otherwise the starting point for the asymptotic expansion is to demand that the condition
\[
	|\Delta_{ij}| \ll |2q_i\cdot k_{ji} + \Gamma_{ij}| 
	\label{eq:condition_convergence}
\]
be fulfilled for the whole range of the loop integration space except for potentially small regions around physical divergences. The distinctive feature of LTD is that since dual propagators only appear in integrands where one loop momentum has been set on-shell, the condition has to be fulfilled in the Euclidean space of the loop three-momemtum. 
Whenever it is satisfied, the dual propagator can be expanded as
\[
	G_D \left( q_i ; q_j\right)&= \sum_{n=0}^\infty \frac{\left(-\Delta_{ij} \right)^n}{\left(2q_i\cdot k_{ji} + \Gamma_{ij} - i0 \eta\cdot k_{ji}\right)^{n+1}}~, 
	\label{eq:expandedpropagator}
\]
or in the case of amplitudes with propagators raised to multiple powers, as often occurs in multiloop amplitudes, by using the generalized binomial theorem
\[
    	\left( G_D \left( q_i ; q_j\right) \right)^m &= \sum_{n=0}^\infty \binom{-m}{n}  \frac{\left(\Delta_{ij} \right)^n}{\left(2q_i\cdot k_{ji} + \Gamma_{ij} - i0 \eta\cdot k_{ji}\right)^{n+m}}~.
\] 
A special case of the above is the situation when $k_{ji}^2 + m_i^2 - m_j^2$ is much smaller than the scalar product $2q_i\cdot k_{ji}$. Then
we must identify $\Gamma_{ij}=0$ and the expansion above simplifies as follows:
\[
	G_D \left( q_i ; q_j\right)&= \sum_{n=0}^\infty \frac{\left(-\Delta_{ij} \right)^n}{\left(2q_i\cdot k_{ji}\right)^{n+1}}~.\label{eq:expandedpropagatorGammaZero}
\]

The asymptotic expansion of the dual propagators given in \eq{eq:expandedpropagator} is the basis for the examples that will be presented in this work. 
In the following, we will disscuss how to select the functions $\Gamma_{ij}$ and $\Delta_{ij}$ in different kinematical limits. Further simplifications are possible  whenever ${\bf k}_{ji}=0$, which will be considered from here on. In that case, with the change of variables $|\mathbf{q_i}| = m_i/2 \, (x_i - x_i^{-1})$, the denominator of the expanded dual propagator takes an easily integrable form. For the case of $\Gamma_{ij}=0$ the denominator of \eq{eq:expandedpropagatorGammaZero} is given by 
\[
 	2q_i\cdot k_{ji} = k_{ji,0} \, m_i \, \left( x_i+x_i^{-1} \right)~,
\]
while in the general case the denominator of \eq{eq:expandedpropagator} can be written as 
\[
	2q_i\cdot k_{ji} + \Gamma_{ij} - \ii \,\eta\cdot k_{ji} = Q_i^2 \left( x_i + r_{ij}\right)\left( x_i^{-1} + r_{ij} \right)~. 
	\label{eq:expandedpropagator_denominator}
\]
The form found here determines the parameters $\Gamma_{ij}$ and $r_{ij}$ appearing in the expansion to be restricted by the conditions 
\[
	\Gamma_{ij} - \ii \,\eta\cdot k_{ji} = Q_i^2 \left( 1+ r_{ij}^2 \right)~, \qquad r_{ij}= \frac{m_i \, k_{ji,0}}{Q_i^2} -\frac{\ii \, \eta\cdot k_{ji}}{Q_i^2}~,
	\label{eq:condition_coefficients}
\]
assuming $|r_{ij}|\le 1$.
For the class of limits where one hard scale $Q$ is available, we can identify $Q_i^2 = \pm Q^2$, where the sign is determined by 
the sign of the hard scale in the expression $k_{ji}^2+m_i^2-m_j^2$. As will be seen in the examples of the following sections this type of expansion facilitates the analytical integration based on integrals of the form 
\[
   	\int_1^\infty \frac{\mathrm{d} x_i}{x_i (x_i + r_{ij})(x_i^{-1} +r_{ij})} = \frac{\log(r_{ij})}{r_{ij}^2-1}, \qquad ~ |r_{ij}|<1~. 
	\label{eq:exampleintegral}
\]
On top of the relations in \eq{eq:condition_coefficients} additional conditions are to be respected by the expansion parameters. The expansion is to converge both at integrand- and at integral-level and the analytic behaviour of the dual propagator may not be fundamentally changed. This is to mean that for a propagator with a singularity the expansion is to also display that singularity, while the expansion of a non-singular propagator is to be finite in all of the integration domain as well. The infinitesimal imaginary prescription of $r_{ij}$ given in \eq{eq:condition_coefficients} accounts properly for the 
complex prescription of the original dual propagator and therefore of its causal thresholds. 
This corresponds to the argument $r_{ij}$ of the logarithm in \eq{eq:exampleintegral} taking a negative value,
${\rm Re}(r_{ij}) < 0$.

While the scenario described above covers many typical limits, asymptotic expansions at thresholds deserve a special treatment since
all the scales are of the same order and, therefore, a hard scale cannot be clearly identified. Even when approaching the physical threshold 
from below and thus considering a dual propagator without pole on the real axis, its behaviour is still strongly influenced by the threshold singularity. 
In cases like this it is necessary to consider the trajectory of the pole in the non-expanded propagator more carefully, which is determined by 
\[
	x_i = -\frac{k_{ji,0}^2+m_i^2-m_j^2 \pm \lambda^{1/2} (k_{ji,0}^2,m_i^2,m_j^2)}{2 k_{ji,0} m_i}~,
	\label{eq:prop_divergence_position}
\]
in terms of the modified K\"all\'{e}n function $\lambda(k_{ji,0}^2,m_i^2,m_j^2)=(k_{ji,0}^2 - (m_i+m_j)^2)(k_{ji,0}^2 - (m_j-m_j)^2 - \ii k_{ji,0}(k_{ji,0}^2+m_i^2-m_j^2)$.
Then, by expanding close to threshold
\[
        \left. x_i \right|_{\beta\to 0^\pm} = - {\rm sign}(k_{ji,0}) \left( 1 \pm \sqrt{- \frac{m_j \, \beta}{m_i} -  \ii k_{ji,0}} + {\cal O}(\beta) \right)~,
        \label{eq:aproxx}
\]
with $k_{ji,0}^2 = (m_i+m_j)^2 (1-\beta)$. Following this procedure we can deduce the correct $r_{ij}$ parameters for the asymptotic expansion 
both from above and from below threshold and bring the dual propagator in the desired form, \eq{eq:expandedpropagator_denominator},
while showcasing the same threshold behaviour as the non-expanded propagator.

In the following sections we will apply these general ideas to benchmark one-loop integrals, and will 
present their asymptotic expansions in several kinematical limits within the LTD formalism.

\section{Asymptotic expansion of the scalar two-point function with two internal masses}
\label{sec:bubble}
 
 \begin{figure}[t]
    \centering
    \begin{fmffile}{bubble}
    \begin{fmfgraph*}(200,200)
        \fmfleft{i}
        \fmfright{o}
        \fmf{plain,tension=5,label=$p$}{i,v1}
        \fmf{plain,tension=5}{v2,o}
        \fmf{plain,left,label=$q_2=\ell,, m$,label.dist=5mm}{v1,v2}
        \fmf{dbl_plain,left,label=$q_1=\ell-p,, M$,label.dist=5mm}{v2,v1}
        \fmfdot{v1,v2}
    \end{fmfgraph*}
    \end{fmffile}
    \caption{The scalar two-point function with scalar particles of masses $M>m$ in the loop.}
    \label{fig:bubble}
\end{figure}
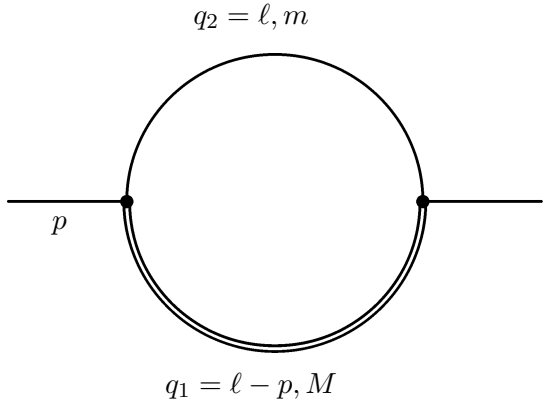

An obvious first benchmark application is the asymptotic expansion of the scalar two-point
function with two different internal masses corresponding to the diagram in figure \ref{fig:bubble}. The corresponding amplitude in the Feynman representation is
\[
	\mathcal{A}^{(1)} = \int_{\ell}  G_F(q_1; M) \, G_F(q_2; m)~, \qquad q_1=\ell-p~, \qquad q_2=\ell~. 
	\label{eq:amplitude_bubble}
\]
The momentum flow, assuming $p_0>0$,  
has been chosen to avoid the appearance of non-causal or unphysical singularities
thus rendering the calculation simpler. These type of integrand singularities, if they appear, always cancel in the sum of dual amplitudes.

We are indeed interested in the asymptotic expansion of the renormalized amplitude, which is well defined directly in four space-time dimensions
\[
	\mathcal{A}^{(1,R)} = \left. \mathcal{A}^{(1)} - \mathcal{A}^{(1)}_\uv \right|_{d=4}~,
	\label{eq:amplitude_bubble_renor} 
\]
where $\mathcal{A}^{(1)}_\uv$ is a local UV counterterm that suppresses the singular behaviour of the unintegrated amplitude for large loop momenta. 
Its Feynman and dual representations are respectively given by 
\[
	\mathcal{A}^{(1)}_\uv &= \int_\ell \left(G_F(\ell; \mu_\uv)\right)^2 = \int_\ell~ \frac{\td{\ell;\mu_\uv}}{2 \left( \ell_{0,\uv}^{(+)}\right)^2}~,
	\qquad  \ell_{0,\uv}^{(+)} = \sqrt{{\boldsymbol \ell}^2 + \mu_\uv^2}~,
	\label{eq:counterterm}
\]
where $\mu_\uv$ is an arbitrary scale. Its integrated form takes the shape 
\[
\mathcal{A}^{(1)}_\uv = \frac{\Gamma(1+\epsilon)}{(4\pi)^{2-\epsilon}} \, \frac{1}{\epsilon} \, \left( \frac{\mu_\uv^2}{\mu^2}\right)^{-\epsilon}~,
\]
and implements the standard $\overline{MS}$ renormalization scheme when identifying the parameter $\mu_\uv$ with the DREG renormalization scale $\mu$. 
The full analytic expression of the renormalized amplitude is well known 
through standard techniques
\[
\mathcal{A}^{(1,R)} &= \frac{1}{16\pi^2} \Bigg[ 2 + \frac{p^2+M^2-m^2}{2p^2} \ln{\frac{\mu_\uv^2}{M^2}} 
+ \frac{p^2+m^2-M^2}{2p^2} \ln{\frac{\mu_\uv^2}{m^2}} \nn \\
&+  \frac{\lambda^{1/2} \left(p^2, m^2,M^2\right)}{p^2} \, \ln{\frac{m^2+M^2-p^2+ \lambda^{1/2} \left(p^2,m^2,M^2\right) }{2mM}}  
 \Bigg]~,
 \label{eq:fulltwo}
\]
which is symmetric under the exchange $m\leftrightarrow M$. This expression will be used to check the validity of the asymptotic expansions
presented in the next sections.

\subsection{Master asymptotic expansion}

The dual representation of the renormalized scalar two-point function (\eq{eq:amplitude_bubble_renor}) is given by
\[
	\mathcal{A}^{(1,R)} = - \int_{\ell} \left[  \td{q_1;M} \, G_D(q_1; \ell) + \td{\ell;m} \, G_D(\ell; q_1)   
	+ \frac{1}{2} \, \td{\ell;\mu_\uv}  \,\left( \ell_{0,\uv}^{(+)}\right)^{-2}\right]~, \label{eq:amplitude_bubble_LTD}
\]
where the dual propagators are
\[
	G_D(q_1; \ell) = \frac{1}{2q_1\cdot p + p^2 - m^2 + M^2 - \ii \, p_0}~, \\
	G_D(\ell; q_1) = \frac{1}{-2\ell\cdot p + p^2 + m^2 - M^2 + \ii \,p_0}~. 
\]
Setting $p=(p_0,\bf{0})$ with $p_0>0$, the on-shell energies and scalar products are $q_{1,0}^{(+)} = \sqrt{\boldsymbol{\ell}^2+M^2}$,  
$\ell_{0}^{(+)} = \sqrt{\boldsymbol{\ell}^2+m^2}$, $q_1\cdot p = q_{1,0}^{(+)} p_0$ and $\ell\cdot p = \ell_{0}^{(+)} p_0$.
With this choice of the reference frame the dual representation and its asymptotic expansion become particularly simple 
as the angular integration of the loop three-momentum is straight. The renormalized result can be reproduced through direct integration of \eq{eq:amplitude_bubble_LTD}.

Still, the general propagator expansion of \eq{eq:expandedpropagator} and \eq{eq:expandedpropagator_denominator} 
can be applied to this amplitude to simplify the integrand
\[
	\mathcal{A}^{(1,R)} &=  - \frac{1}{16\pi^2} 
	\Bigg[  \sum_{i,j=1,2} \frac{m_i^2}{Q_i^2} \sum_{n=0}^\infty \left(-\frac{\Delta_{ij}}{Q_i^2}\right)^n \, I^{(n)} (r_{ij}, m_i) 
	 + I_\uv (\mu_\uv) \Bigg]~, 
	 \label{eq:A_solution_unrenormalized}
\]
where $m_1=M$, $m_2=m$ and the remaining integrals are contained in 
\[
	I^{(n)} (r_{ij}, m_i) = \lim_{\Lambda\to \infty}
	\int_1^{\frac{\Lambda+\sqrt{\Lambda^2+m_i^2}}{m_i}} \text{d} x \, \frac{(x^2-1)^2 \, x^{-3}} { \left[ \left( x + r_{ij} \right) \left(x^{-1} + r_{ij} \right)\right]^{n+1}}~,
\]
and
\[
	I_\uv (\mu_\uv) = \lim_{\Lambda\to \infty}\int_1^{\frac{\Lambda+\sqrt{\Lambda^2+\mu_\uv^2}}{\mu_\uv}} \text{d} x \, \frac{2 (x^2-1)^2 \, x^{-1}} {\left(x^2+1\right)^2}~. 
\]
We have introduced a cutoff $\Lambda$ because the individual contributions are still singular in the UV. The sum of all of them 
is UV finite, however. Therefore, we can safely work in four space-time dimensions and then take the limit $\Lambda\to \infty$ after integration. 
Notice that the cutoff is a valid regulator because it acts on the Euclidean space of the loop three-momentum. 
The results of these integrals, up to order $n=2$, are given by
\[
	I^{(n)} (r_{ij}, m_i)   
	&\overset{n=0}{=} \lim_{\Lambda\to \infty} \left[ \frac{2\Lambda}{m_i\, r_{ij}} - \left(1+\frac{1}{r_{ij}^2} \right) \ln{\frac{2\Lambda}{m_i}} 
	+ \left( 1 - \frac{1}{r_{ij}^2} \right) \ln{r_{ij}} \right]~, \nn \\ 
	&\overset{n=1}{=} \lim_{\Lambda\to \infty} \left[-\frac{1}{r_{ij}^2}  \left( 1 - \ln{\frac{2\Lambda}{m_i}} - \frac{1+r_{ij}^2}{1-r_{ij}^2} \ln{r_{ij}} \right) \right]~, \nn \\
	&\overset{n=2}{=} \frac{1}{\left( 1-r_{ij}^2 \right)^2} 
	\left( \frac{1+r_{ij}^2}{2r_{ij}^2} + \frac{2}{1-r_{ij}^2} \ln{r_{ij}}    \right)~,
	\label{eq:I_solution}
\]
and
\[
	I_\uv (\mu_\uv) = \lim_{\Lambda\to \infty} \left[ 2 \ln{\frac{2\Lambda}{\mu_\uv}} - 2\right]~.
\]
A noteworthy feature of this expansion is that the UV divergence  lessens with each order in the expansion. 
Indeed, all the contributions with $n\ge 2$ are UV finite, and can be calculated directly by extending the upper limit of the integral to infinity.  
The linearly UV divergent terms appearing at $n=0$ cancel between the two 
dual amplitudes and the logarithmic dependence on the UV cutoff $\Lambda$ of both terms at $n=0$ and $n=1$ is canceled by the UV counterterm. 

The asymptotic expansion of the renormalized amplitude takes the general form
\[
	\mathcal{A}^{(1,R)} = & \frac{1}{16\pi^2 } \sum_{i,j} \left[2 + c_{0,i} \ln{\frac{\mu_\uv}{m_i}}
	+ \sum_{n=0}^\infty \left( c^{(n)}_{1,i} + c^{(n)}_{2,i}  \ln{r_{ij}} \right)  \right]~.
	\label{eq:asymptotic_general}
\]
The coefficient $c_{i,0}$ is given by 
\[
c_{0,i} = \frac{m_i^2}{Q_i^2} \left( 1 + \frac{1}{r_{ij}^2}\left( 1 + \frac{\Delta_{ij}}{Q_i^2}\right) \right) =\frac{p^2+m_i^2-m_j^2}{p^2}~,
\]
and the coefficients $c^{(n)}_{1,i}$ and $c^{(n)}_{2,i}$ needed for the first few orders of the expansion are given by 
\[
  c^{(n)}_{1,i} &= - \frac{m_i^2}{Q_i^2} \left\{ 0~ ,~ \frac{-\Delta_{ij}}{Q_i^2} \frac{-1}{r_{ij}^2 }~ ,~ \left( \frac{-\Delta_{ij}}{Q_i^2}\right)^2  \frac{1+r_{ij}^2}{2r_{ij}^2(1-r_{ij}^2)^2}~ ,~ \left( \frac{-\Delta_{ij}}{Q_i^2}\right)^3 \frac{1 +10r_{ij}^2+r_{ij}^4}{6r_{ij}^2(1-r_{ij}^2)^4} \right\}~, \nn \\
  c^{(n)}_{2,i} &= - \frac{m_i^2}{Q_i^2} \left\{ 1- \frac{1}{r_{ij}^2}~ ,~ \frac{-\Delta_{ij}}{Q_i^2} \frac{1+r_{ij}^2}{r_{ij}^2 (1-r_{ij}^2)}~ ,~ \left( \frac{-\Delta_{ij}}{Q_i^2}\right)^2  \frac{2}{(1-r_{ij}^2)^3}~ ,~ \left( \frac{-\Delta_{ij}}{Q_i^2}\right)^3 \frac{2(1+r_{ij}^2)}{(1-r_{ij}^2)^5} \right\}~. 
 \label{eq:cis} 
\]
Each term of the expansion is suppressed by extra powers of $\Delta_{ij}$.

\subsection{Asymptotic expansion for different kinematical limits}

We now consider explicitly different kinematical limits and the corresponding asymptotic expansions.
In the limit of one large mass, $M^2\gg \{m^2,p^2\}$, the expansion parameters are 
$Q_1^2 = - Q_2^2 = M^2$, $r_{12} =\sqrt{p^2}/M$, and $r_{21} = m \sqrt{p^2}/M^2$. 
The functions $\Gamma_{ij}$ and $\Delta_{ij}$ are summarized in table~\ref{tab:coefficients_bubble}.

The election of the expansion parameters in the limit of a large external momentum, $p^2\gg \{m^2,M^2\}$, is also summarized in table~\ref{tab:coefficients_bubble}. Since this kinematical configuration is above threshold, the asymptotic expansion should feature an imaginary part just as the original integral. This imaginary part is generated through $\ln{r_{21}}$, with $r_{21} = - m/\sqrt{p^2}+\ii$.

\begin{figure}
	\centering
	\includegraphics[width=0.5\textwidth]{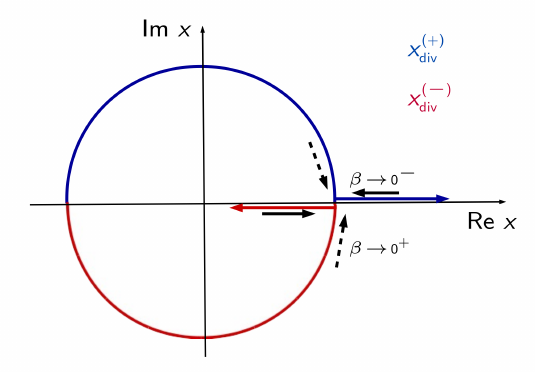}
	\caption{The position of the singularities of the unexpanded dual propagator in terms of the threshold parameter $\beta = 1 - \frac{p^2}{(m+M)^2}$ after the change of variable . The path of integration goes from the point where the singularities reach the real axis until infinity.}
	\label{fig:polposition}
\end{figure}

In the slightly more involved case of the threshold limit with $\beta= 1-p^2/(m+M)^2 \to 0^\pm$, 
the election of the expansion parameters, also summarized in table~\ref{tab:coefficients_bubble},
is such that $\log(r_{12}) = \sqrt{m/M} \sqrt{-\beta-\ii}$ and $\log(r_{21}) = \sqrt{M/m} \sqrt{-\beta+\ii}- \imath \pi$. 
This compact result is obtained by exponentiating the expanded expression determining the position of the threshold in 
the complex plane given by \eq{eq:aproxx}.
The expressions for $\Delta_{ij}$ and $Q_i^2$ fulfill the necessary asymptotic behaviour as  
\[
\frac{m_i^2}{Q_i^2} = \frac{m_i}{m+M} + {\cal O}(\beta^{1/2})~, \qquad
\frac{\Delta_{ij}}{Q_i^2} = {\cal O}(\beta^2)~. 
\]
The first dual propagator $G_D(q_1;\ell)$ is free of threshold singularities and thus leads to 
a real expansion independently of the sign of $\beta$. The expressions obtained for both $r_{12}$ and $r_{21}$ can be used both when 
approaching the threshold from below and from above because the infinitesimal imaginary component accompanying  $\beta$
is fixed by the complex prescription of the dual propagators.
While in the case below threshold no singularity appears in the propagator on the real axis, it does exist in the complex plane and approaches 
the path of integration as can be seen in figure \ref{fig:polposition}. 

In all the kinematical regions studied, we have achieved their asymptotic expansions by conveniently selecting the  
expansion parameters that are used in a common expression, \eq{eq:asymptotic_general}, that describes all these limits
at once. In each limit fast convergence was achieved both at integrand- and at integral-level.

\begin{table}
\begin{center}
\begin{tabular}{c|c|c|c} \hline 
  & $M^2 \gg \{m^2, p^2\}$ & $p^2\gg \{m^2, M^2\}$  & $p^2= (m+M)^2 (1-\beta)~, \quad  \beta\to 0^\pm$ \\ \hline \hline
 $G_D(q_1;\ell)$  & & & \\ 
                             $\Gamma_{12}$ & $M^2+p^2$ & $p^2+M^2$ &  $2 M p \, \cosh\left(\sqrt{-\frac{m \, \beta}{M}-\ii}\right)$ \\
                             $\Delta_{12}$     & $-m^2$ & $-m^2$ & $p^2+M^2-m^2 - \Gamma_{12}$  \\
                             $r_{12}$ & $\frac{\sqrt{p^2}}{M}$ & $\frac{M}{\sqrt{p^2}}$ & $\exp\left(\sqrt{-\frac{m\, \beta}{M}-\ii}\right)$  \\
                             $Q_1^2$    & $M^2$  & $p^2$ & $M \, p \, \exp\left(- \sqrt{-\frac{m \, \beta}{M}-\ii}\right)$    \\ \hline
 $G_D(\ell; q_1)$ & & & \\ 
                             $\Gamma_{21}$ & $- M^2  - \frac{m^2\, p^2}{M^2}$& $p^2+m^2$ & $2 m \, p \, \cosh\left(\sqrt{-\frac{M \, \beta}{m}+\ii}\right)$ \\
                             $\Delta_{21}$     & $p^2+m^2+\frac{m^2 \,p^2}{M^2}$ & $-M^2$ & $p^2+m^2-M^2-\Gamma_{21}$ \\
                             $r_{21}$ & $\frac{m\sqrt{p^2}}{M^2}$ & $-\frac{m}{\sqrt{p^2}} + \ii$ & $- \exp\left(\sqrt{-\frac{M\, \beta}{m}+\ii}\right)$ \\
                             $Q_2^2$    & $-M^2$  & $p^2$ & $m \, p \, \exp\left(- \sqrt{-\frac{M \,\beta}{m} +\ii}\right)$   \\ \hline
\end{tabular}
\caption{The coefficients of the asymptotic expansions for the scalar two-point function.}
\label{tab:coefficients_bubble}
\end{center}
\end{table}

\subsection{Comparison with \textit{Expansion by Regions}}

It is of interest to see how the expansions developed above hold up in comparison to the established method of Expansion by Regions (EbR)~\cite{Beneke:1997zp, Smirnov:2002pj,Pak:2010pt,Jantzen:2011nz,Jantzen:2012mw,Mishima:2018olh,Semenova:2018cwy}. Within this successful method the integrand of the Feynman amplitude, written in terms of Minkowski momenta, is expanded by dividing the space of the loop momenta into distinct regions. In each region, the integrand is expanded into a Taylor series with respect to the parameters considered small therein. Consecutively, the expanded integrands are integrated over the whole integration domain, not just within the region where the expansion was justified. The scaleless integrals that may appear are set to zero as commonly done within DREG. While Expansion by Regions has been successful for many types of amplitudes a general proof is still pending. 
One may raise a few issues with the procedure above which will be mentioned in the context of its application to the scalar two-point function in \eq{eq:twopointfeynman}.
We centre the discussion in the limit of one large mass, $M^2\gg \{m^2,p^2\}$.

While in a general loop integral many types of regions can appear in this particular example there are only two regions, the hard region with $\ell \sim M$ and the soft region with $\ell \sim \{ m, \sqrt{p^2} \}$ \footnote{To be precise,  the scaling is assumed for  the momentum in the \textit{Euclidean sense}, i.e. $|\ell| = \sqrt{\ell_0^2 + \boldsymbol{\ell}^2}$.}. The scalar product between the loop momentum and the external momentum inherits the scaling of the momentum itself, that is for the hard region one performs the replacements
\[
	\{\ell^2, M^2\} \to \lambda^2 ~\{\ell^2, M^2\}~, \qquad p\cdot\ell \to \lambda ~ p\cdot \ell~,
\]
and expands for $\lambda\to\infty$. 
The assumed relationship between the large loop momentum and both its square and its scalar products does not account for cancellations between the energy component and the spatial components which will take place when integrating over the unrestricted components of the loop momentum. The Taylor series with the prescriptions above and comparable ones for the soft region leads to the expanded integrands
\[
	\mathcal{A}^{(1)}_\text{hard} &= \int_\ell \left( \frac{1}{\ell^2} + \frac{m^2}{(\ell^2)^2}+ \frac{m^4}{(\ell^2)^3} + \dots \right) \left( \frac{1}{\ell^2-M^2} +\frac{2p\cdot \ell - p^2}{(\ell ^2-M^2)^2}+\frac{4(p\cdot \ell)^2 + p^2}{(\ell ^2-M^2)^3} + \dots\right)~, \nonumber\\
	\mathcal{A}^{(1)}_\text{soft} &= \int_\ell \frac{1}{\ell^2-m^2} \left( - \frac{1}{M^2} - \frac{(p-\ell)^2}{M^4} - \frac{(p-\ell)^4}{M^6} + \dots \right)~.
\]

The first integrated order of the expansion is achieved by combining the counterterm with the first term appearing in the hard region
\[
	\int_\ell  \frac{1}{\ell^2(\ell^2-M^2)} - \mathcal{A}_\uv^{(1)}  = \frac{1-\ln{\frac{M^2}{\mu_\uv^2}} }{16\pi^2} + {\cal O}(\epsilon)~.
\]
The soft region does not contribute at this order. For the next order one must select all terms in the expansion at integrand-level which will lead to contributions of order $M^{-2}$. This includes the first term of the expansion in the soft region and three terms from the hard region. The result achieved in this way is indeed the Taylor series ($\mathcal{T}$) of the full result
\[
	\mathcal{T} \mathcal{A}^{(1,R)} \left( M, \infty \right) =  \frac{1 }{16\pi^2} \left( 1- \ln{\frac{M^2}{\mu_\uv^2}} 
	+ \frac{p^2 - 2 m^2 \ln{\frac{M^2}{m^2}}}{2 M^2}  + \dots \right)~.
	\label{eq:taylor}
\]

In direct comparison, we give here the first renormalized orders of the series achieved through the general expansion of 
the dual propagator \eq{eq:asymptotic_general} in the limit of one large mass, with reorganized logarithms:
\[
	\mathcal{A}^{(1,R)}_{n\le1} &= \frac{1}{16 \pi ^2} \Bigg(1- \ln{\frac{M^2}{\mu_\uv^2}}-\frac{m^2}{M^2}
	-  \frac{m^2 ( M^2+m^2+p^2 )}{M^4 - m^2 p^2} \ln{\frac{M^2}{m^2}}  \nn \\
 	&+ \frac{m^2 ((p^2)^2-m^2 M^2)}{(M^2-p^2) (M^4-m^2 p^2)} \ln{\frac{M^2}{p^2}} \Bigg)~. 
        	\label{eq:ltdexpansion}
\] 
By including the next term of the expansion ($n=2$) and then expanding the rational coefficients for $M^2\gg \{m^2, p^2\}$,
we recover the expected Taylor series. Higher terms of the Taylor series can be obtained by including more terms in the dual expansion, 
$n\geq~3$. The asymptotic expansions in \eq{eq:taylor} and \eq{eq:ltdexpansion}
display the same logarithmic dependence, although they differ in the rational coefficients accompanying the logs, which 
partially encode higher orders in the expansion. This is due to the fact that $\Delta_{12}$ includes subleading terms. 
The expression in \eq{eq:ltdexpansion} also contains
a logarithmic dependence on $\ln{M^2/p^2}$, which is formally one order higher than \eq{eq:taylor}
and cancels when more terms in the dual expansion are included.

\begin{table}
\begin{center}
\begin{tabular}{c|c c c||c c} 
  & $G_D$ expansion & $\mathcal{O}(M^{-2})$ & $\mathcal{O}(M^{-4})$ & & EbR \\ \hline 
 $n=1$ & $2.67\%$ & $2.45\%$ & $2.68\%$ & $\mathcal{O}(M^{-0})$ & $3.34\%$ \\
 $n=2$ & $0.0375\%$ & $0.135\%$ & $0.0300\%$ & $\mathcal{O}(M^{-2})$ & $0.135\%$ \\
 $n=3$ & $7.60\cdot 10^{-6}$ & $0.135\%$ & $6.18 \cdot 10^{-5}$ & $\mathcal{O}(M^{-4})$ & $6.18 \cdot 10^{-5}$ 
\end{tabular}
\caption{The relative errors with respect to the full result of the dual expansion as given in \eqref{eq:asymptotic_general}
(both the direct result of this expansion and considering only the leading behaviour at large $M$) compared with the result obtained through Expansion by Regions. Evaluation with parameters $M=10m$, $p^2=3m^2$, and $\mu=M$.}
\label{tab:comparison_dualexp_EbR}
\end{center}
\end{table}

The relative error obtained by the two expansions with respect to the full result (\ref{eq:fulltwo}) is numerically of the same order of magnitude. For the values of $M=10m$, $p^2=3m^2$ and $\mu_\uv=M$ the relative error obtained including only the leading term in EbR is $3.3\%$ compared to the $2.7\%$ obtained by expanding the dual propagator as described above. Including one more order in the expansion the results are given by $0.14\%$ and $0.038\%$, respectively. Interestingly, the numerical error at leading order of the expansion of the dual propagators can be reduced by expanding the rational coefficients. The comparison of the results obtained in EbR  with those based on the expansion of the dual propagator is summarized in table~\ref{tab:comparison_dualexp_EbR}.

There is a distinction in the application of the two methods which we would like to emphasize. In EbR it is essential to consider the terms of the expansion at integrand-level to pick out only those which will contribute at a given order of the result. Failing to do so does not only lead to numerical differences but will generally lead to divergent results. This is due to the cancellation between UV and IR singularities appearing in the expansions of the soft and hard region. In contrast, UV renormalization within the method of expanding the dual propagators only involves the lowest orders of the integrand-level expansion. Including higher terms is optional for improving numerical precision and for this purpose it is possible to straight-forwardly include any amount of terms without needing to worry about ensuring cancellations between separate regions.

\subsection{Asymptotic expansion by dual regions}

\begin{figure}
	\centering
	\includegraphics[width=0.5\textwidth]{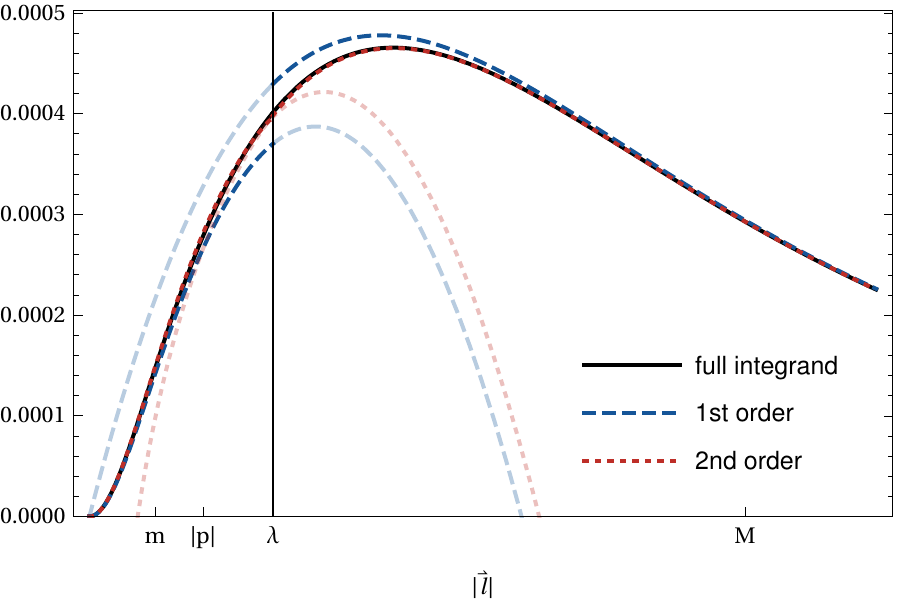}
	\caption{The convergence at integrand-level of the expansion given in \eq{eq:Taylor_expansion} for the values $M=10\, m$, $p^2=3\, m^2$, and $\mu_\uv=M$. \label{fig:Taylor_integrand_convergence}}
\end{figure}

The properties of dual amplitudes can also be exploited in a more direct way to facilitate asymptotic expansions. After applying LTD to the integrand of a Feynman integral the loop momentum is restricted to Euclidean on-shell values. It is thus possible to directly expand the integrand into a Taylor series with respect to whichever scale is considered to be small. These expansions can be done anywhere within the integration domain and depend on the size of the loop three-momentum. In the example of the scalar two-point function in the limit of one large mass a region with small and one with large loop three-momentum are to be distinguished, which we chose to call \textit{dual regions} since they become accessible only after obtaining a Euclidean integration domain through the application of LTD. For any loop three-momentum appearing during the integration one of these expansions is well justified and convergent such that the integration domain can be split up into well-defined integrand-level expansions as
\[
 \mathcal{A}^{(1,R)} =  \int_0^\infty \mathrm{d} |\boldsymbol{\ell}| ~ a (\boldsymbol{\ell}) = \int_0^\lambda\mathrm{d} |\boldsymbol{\ell}|~ \mathcal{T} a (M,\infty) + \int_\lambda^\infty\mathrm{d} |\boldsymbol{\ell}|~ \mathcal{T} a  (\{ \boldsymbol{\ell},M\}, \infty)~. \label{eq:Taylor_expansion}
\]
The integrand-level convergence and the behaviour around the matching point where the expansions change can be seen in figure~\ref{fig:Taylor_integrand_convergence}.  

\begin{figure}
	\centering
	\includegraphics[height=0.27\textwidth]{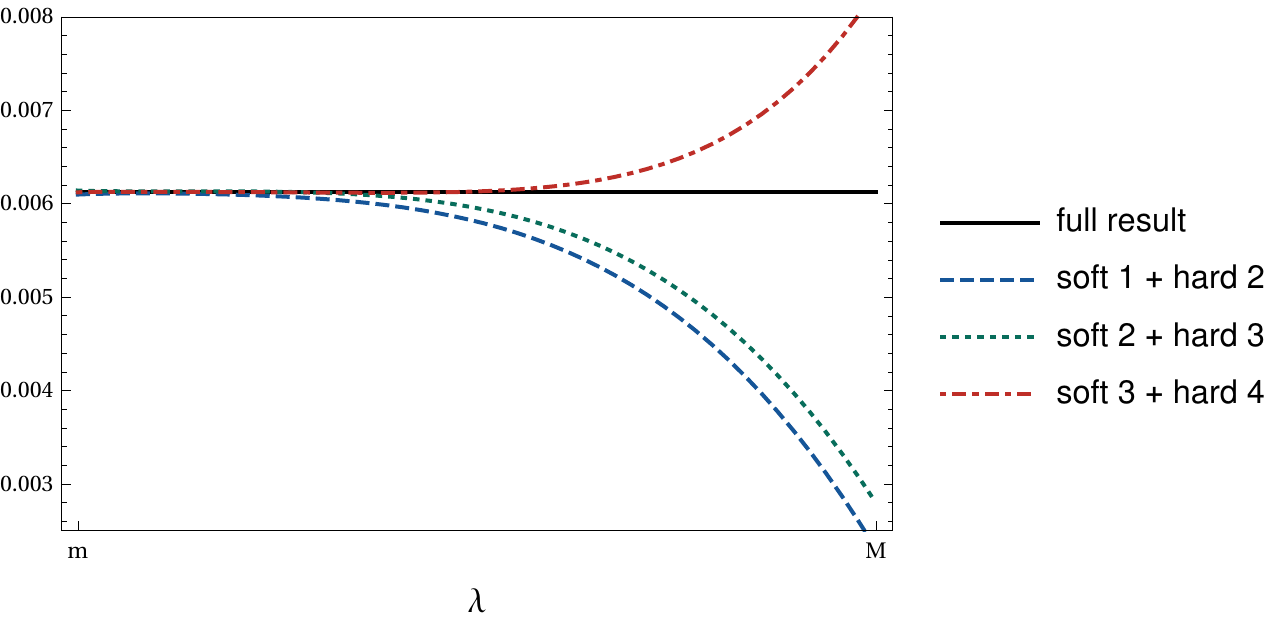}
	\includegraphics[height=0.27\textwidth]{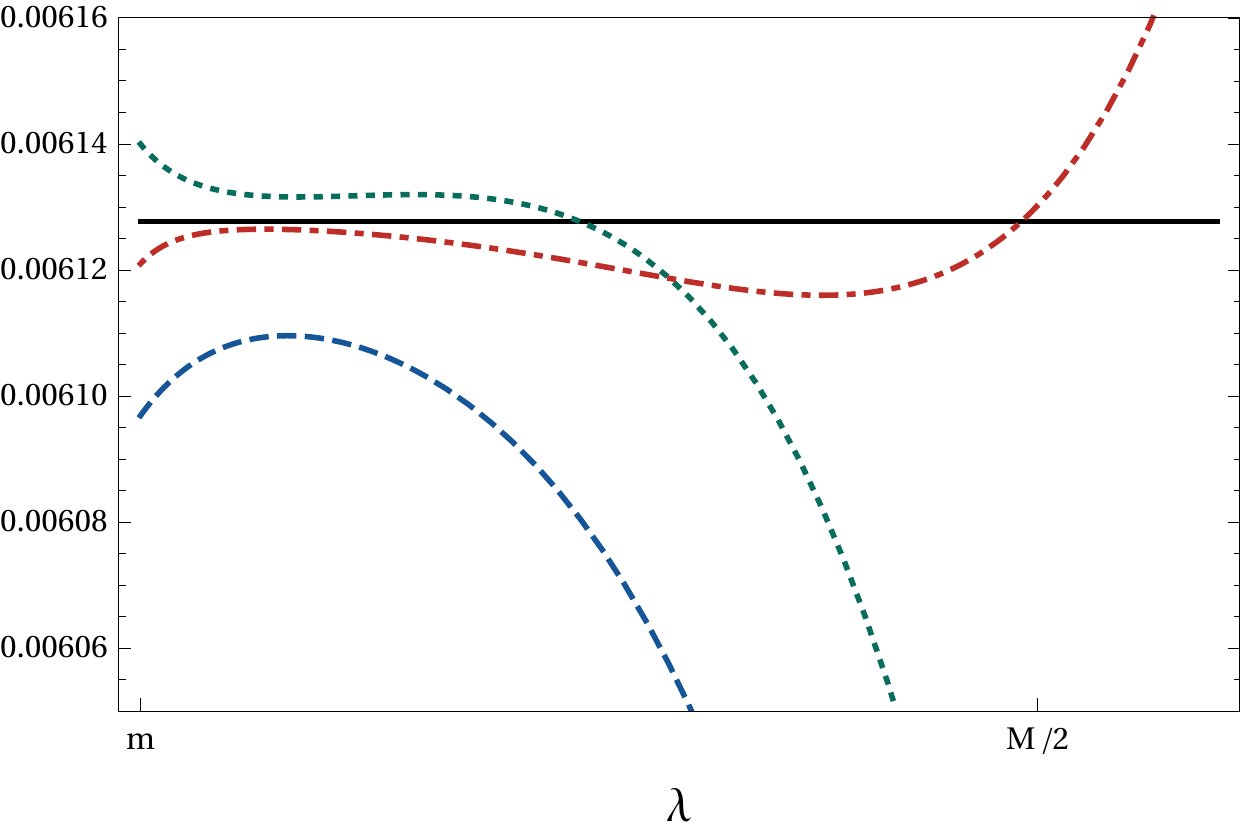}
	\caption{The convergence of the integrated results of the expansion given in \eq{eq:Taylor_expansion} for the values $M=10\, m$, $p^2=3\,m^2$, and $\mu_\uv=M$, in the range of $m<\lambda<M$, with the image on the right showing a closeup of the region where the optimal value of $\lambda$ can be identified.}	
	\label{fig:Taylor_integral_convergence}
\end{figure}

Integration of this type of expansion is straight-forward, as the integrand gets quite simple. Including only the first order of the Taylor expansion in the hard region one obtains the result
\[
	\mathcal{A}^{(1,R)}_{\rm hard} = \int_\lambda^\infty\mathrm{d} |\boldsymbol{\ell}|~ \mathcal{T}_0 ~ a  (\{ \boldsymbol{\ell},M\}, \infty) &= \frac{1}{16\pi^2} \left( 1-\frac{2 \lambda ^2 }{M^2} \left(1-\frac{1}{\sqrt{1+\frac{M^2}{\lambda ^2}}}\right) \right)~,
\]
which with the choice of $\lambda=M$ reduces to
\[	
	\left. \mathcal{A}^{(1,R)}_{\rm hard} \right|_{\lambda\to M} &= \frac{\sqrt{2}-1}{16\pi^2}~.
\]
While in principle the choice for the order of integrand-level expansion in the soft and the hard region is independent of each other, and renormalization is guaranteed in any case, a comparable accuracy in both regions can be expected only when combining the $(n+1)$-th term of the hard region with the $n$-th term of the soft region. Depending on the value of the parameter $\lambda=m+x (M-m)$ these results are shown in figure \ref{fig:Taylor_integral_convergence}. We have order-to-order convergence at the integrated level of the expansion when considering both the case $\lambda=m$ and $\lambda=M$. While the accuracy of the result does increase with each order, the overall relative error remains comparatively large as shown in table \ref{tab:Taylor_integral_convergence_edges}. A higher level of precision can be obtained by evaluating the results at those values for $\lambda$ for which local extrema appear. 

\begin{table}
\begin{center}
\begin{tabular}{c|c c|c c | c c} 
 &  $\lambda=m$: & & $\lambda=M$: & &$\lambda$ optimized\\
 & result & rel. error& result & rel. error & result & rel. error \\ \hline 
 soft 1 and hard 2 & $0.006097$ & $0.5\%$ & $0.002302$ & $91\%$ & $0.006110$ & $0.3\%$\\
 soft 2 and hard 3 & $0.006140$ & $0.2\%$ & $0.002796$ & $75\%$ & $0.006132$ & $0.06\%$ \\
 soft 3 and hard 4 & $0.006121$ & $0.1\%$ & $0.008284$ & $30\%$ & $0.006126$ & $0.02\%$
\end{tabular}
\caption{Integrated results of the expansion in \eq{eq:Taylor_expansion} evaluated at $\lambda=m$ and $\lambda=M$, with parameters $M=10m$, $p^2=3m^2$, and $\mu=M$, compared to the full result of $0.006128$.}
\label{tab:Taylor_integral_convergence_edges}
\end{center}
\end{table}

\begin{figure}
	\centering
	\includegraphics[height=0.315\textwidth]{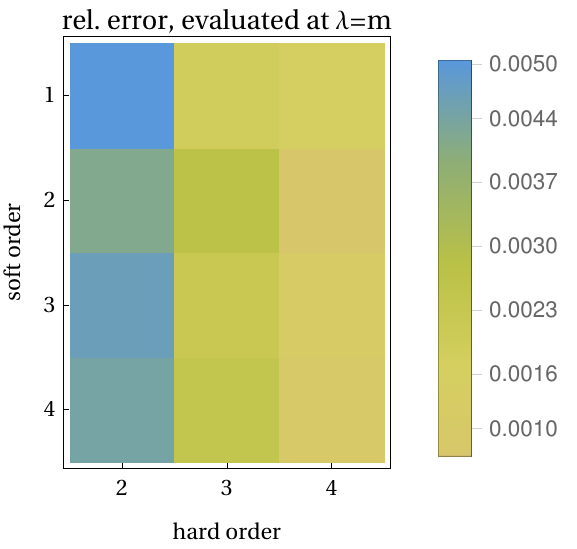}
	\includegraphics[height=0.315\textwidth]{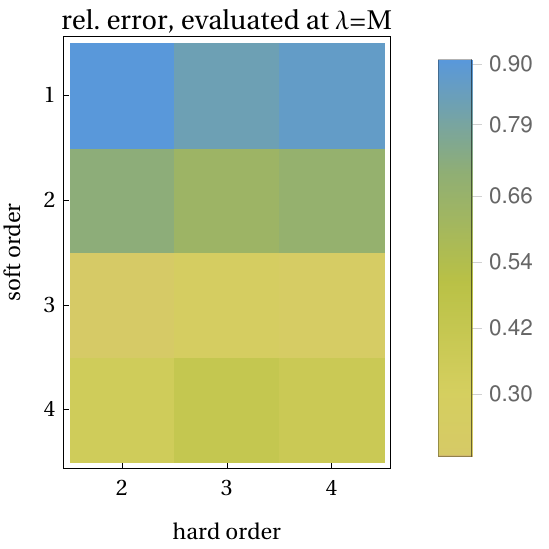}
	\includegraphics[height=0.315\textwidth]{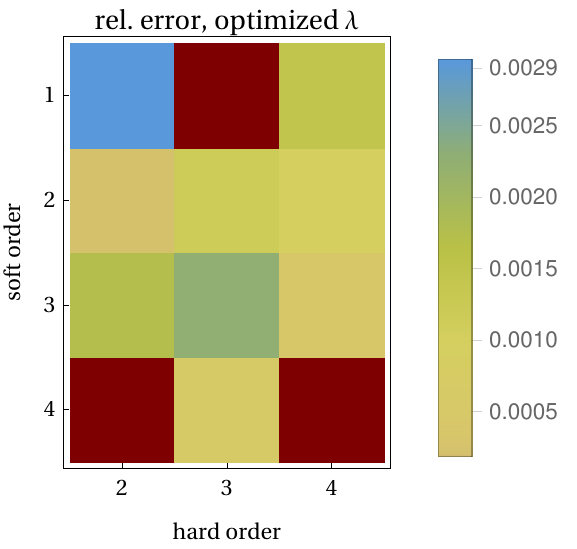}
	\caption{The relative error of the integrated Taylor expansion, comparing the three scenarios of switching between soft and hard region at $\lambda=m$, $\lambda=M$ or by choosing the value of $\lambda$ that corresponds to an extremum. The order of the expansion in the soft (hard) region is increased along the vertical (horizontal) axis. The used values for the scales are $M=10\,m$, $p^2=3\,m^2$, and $\mu_\uv=M$. In three cases the result did not have any extrema thus the relative error could not be optimized as per the method described, recognizable on the far right by the dark squares.
	}
	\label{fig:Taylor_relerror}
\end{figure}

Since adding additional orders to the expansion does not produce any artifical singularities any order of the expansion in the soft region can be combined with any other order in the hard region. This can lead to very precise results already at low orders in the expansion, as can be seen in figure \ref{fig:Taylor_relerror} for the case of order $2$ in both expansion. Clear convergence at the level of the result can be observed for all three scenarios of choosing $\lambda$ when combining expansions of comparable accuracy as explained above.

\begin{figure}
	\centering
	\includegraphics[width=0.8\textwidth]{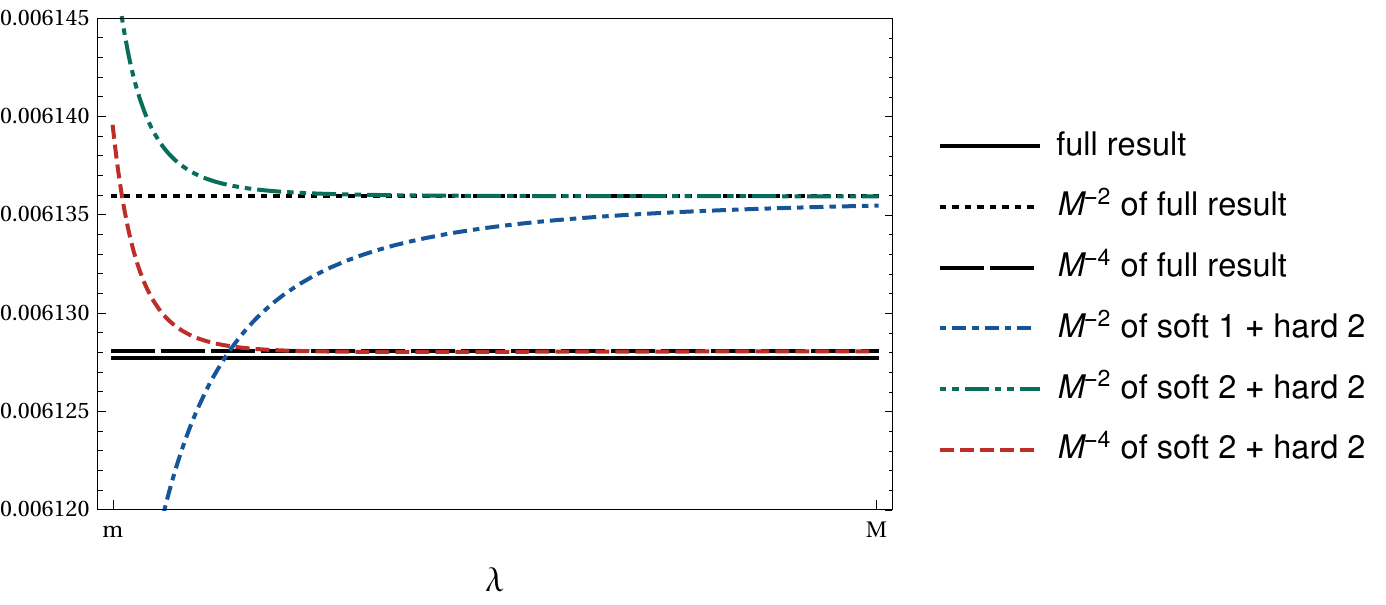}
	\caption{Recovery of the Taylor series of the full result for the scalar two-point function by applying a second expansion to the integrated result of the combined Taylor expansions of the integrand at $M=10\, m$, $p^2=3\, m^2$, and $\mu_\uv=M$.}
	\label{fig:reproduction_Taylor}
\end{figure}

The integrated results obtained by using Taylor expansions at integrand-level contribute not just to one order in $M$. Nonetheless, the Taylor series of the full result, which alternatively has been obtained using EbR, can be recovered using this method. The results as shown in figure \ref{fig:Taylor_integral_convergence}, still depending on an undetermined $\lambda$, may be expanded a second time for $M\to \infty$. As can be seen in figure \ref{fig:reproduction_Taylor} the Taylor series of the full result is then approximated by setting $\lambda=M$.

The main advantage of this Taylor series inspired expansion method is its easy application and automatization. The behaviour portrayed here has been seen as well in a modification of the scalar-two point function which has been rendered \uv-finite by increasing the power of one of the propagators, with the only relevant difference being that the same order of both the soft and the hard integrand-level expansion had to be combined.

\section{Asymptotic expansion of the scalar three-point function} 
\label{sec:scalar3pt}

\begin{figure}
    \centering
    \begin{fmffile}{triangle}
    \begin{fmfgraph*}(200,200)
		\fmfleft{i1,i2}
		\fmfright{o1}
		\fmf{plain,label=$p_1$,tension=2}{i1,v1}
		\fmf{plain,label=$p_2$,tension=2}{i2,v2}
		\fmf{plain,label=$-p_{12}$,tension=2}{v3,o1}
		\fmf{plain,label=$q_3=\ell,, M$}{v1,v3}
		\fmf{plain,label=$q_2=\ell+p_{12},, M$}{v3,v2}
		\fmf{plain,label=$q_1=\ell+p_1,, M$}{v2,v1}
		\fmfdotn{v}{3}
        \fmfdot{v1,v2,v3}
    \end{fmfgraph*}
    \end{fmffile}
    \caption{The scalar three-point function with scalar particles of equal mass$M$ in the loop.}
    \label{fig:triangle}
\end{figure}
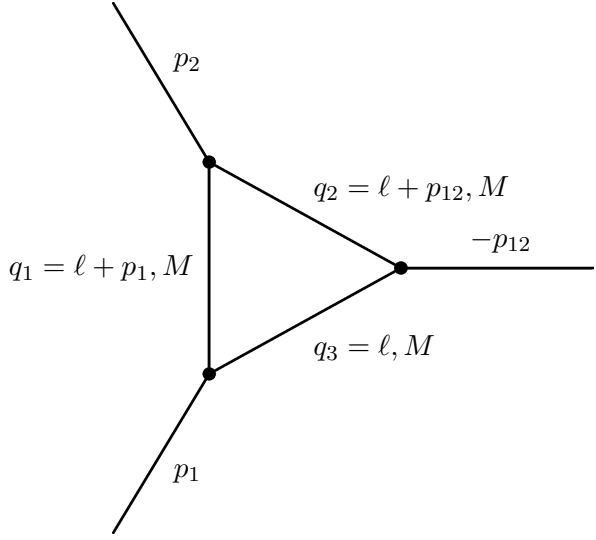

As a benchmark application with more external legs, we consider 
the scalar three-point function at one-loop as shown in figure \ref{fig:triangle} with all the internal masses equal 
\[
	{\cal A}_3^{(1)} = \int_\ell G_F(q_1,q_2,q_3; M)~,
\]
where $G_F(q_1,q_2,q_3; M) = \prod_{i=1}^3 G_F(q_i, M)$, 
with $q_1=\ell+p_1$, $q_2 = \ell+p_1+p_2$ and $q_3=\ell$. 
Applying LTD to this integral leads to
\[
	{\cal A}_3^{(1)} = - \int_\ell \left[ \td{q_1} G_D (q_1; q_2, q_3) + \td{q_2} G_D (q_2; q_1, q_3) + \td{q_3} G_D (q_3; q_1, q_2) \right]~, \label{eq:A3_dual}
\]
with $G_D(q_i; q_j, q_k) = G_D(q_i; q_j) \, G_D(q_i; q_k)$ based on the dual propagators given as in \eq{eq:dualpropagator}. 
The three different linear combinations of external momenta that appear in the dual propagators are
\[
	k_{12} = -k_{21} = -p_2~,\qquad k_{13}=-k_{31} = p_1~,\qquad \text{and} \qquad  k_{23}=-k_{32} = p_1+p_2~.
\]
Only one of these can be chosen to have a vanishing three-momentum, for example by using the center-of-mass system of $p_1$ and $p_2$, therefore ${\bf p}_{12}=0$. The complete dual integrand thus only has angular dependence in the scalar products $q_i\cdot p_1$ and $q_i\cdot p_2$.

In the special case when only one particle with mass $M$ runs through the loop and two of the external particles are massless ($p_1^2=p_2^2=0$ and $p_{12}^2 = s_{12}$) the LTD representation condenses to
\[
	{\cal A}_3^{(1)} &= - \int_\ell \Bigg\{ -\frac{\td{q_1; M}}{2q_1 \cdot p_{12}} \left(\frac{1}{2 q_1\cdot p_1}+ \frac{1}{2 q_1\cdot p_2} \right) \nn \\
	&+ \frac{\td{\ell; M}}{(-2\ell\cdot p_2) (-2\ell\cdot p_{12}+s_{12}+\ii)}
	+ \frac{\td{\ell; M}}{(2\ell\cdot p_1) (2\ell\cdot p_{12}+s_{12})} 
	\Bigg\}~, \label{eq:triangle} 
\]
with the on-shell energies $q_{1,0}^{(+)} = \sqrt{({\lb + \pb_1})^2+M^2}$ and $q_{2,0}^{(+)} = q_{3,0}^{(+)} = \ell_{0}^{(+)}=  \sqrt{\lb^2+M^2}$. Thus, the on-shell energies in the second and third on-shell cuts are identical.

Shifts in the three-momentum are unproblematic as long as the integrand contains neither IR nor UV singularities that require local cancellations between dual amplitudes.  We may therefore use the following identity
\[
\int_\ell \frac{\td{q_1; M}}{(2 q_1\cdot p_{12})(2 q_1\cdot p_i)} = \int_\ell \frac{\td{\ell; M}}{(2 \ell \cdot p_{12})(2\ell \cdot p_i)}~.
\]
Then, the loop three-momentum can be parametrized as
\[
	\mathbf{\ell} = | \mathbf{\ell} | \left( 2 \sqrt{v(1-v)} \hat{e}_\perp , 1-2v \right)~,
\]
where $\hat{e}_\perp$ is the unit vector perpendicular to $\mathbf{p}_1$.
The angular dependence then takes the shape
\[
2 \ell\cdot p_i = \sqrt{s_{12}} \, (\ell_0^{(+)} \mp |{\lb}| (1- 2v) )~, \qquad i=1,2~.
\]
In this expression the two angular integrations are related by the change of variables $v\to 1-v$ and thus
\[
	\int_0^1 \frac{dv}{2\ell\cdot p_1} = \int_0^1 \frac{dv}{2\ell\cdot p_2}=\frac{1}{2\sqrt{s_{12}}\, |\lb|} \, \ln{\frac{\ell_0^{(+)}+|\lb|}{\ell_0^{(+)}-|\lb|}}~,
\]
where the usual change of variables $|\lb| = M/2 \, (x-x^{-1})$ with $x>1$ has been employed. Consequently, we can rewrite \eq{eq:triangle} as
\[
	{\cal A}_3^{(1)} =  \int_\ell \frac{\td{\ell; M}\, s_{12}}{(2\ell\cdot p_{12})(2\ell\cdot p_1)} \Bigg\{ 
	\frac{1}{-2\ell\cdot p_{12}+s_{12}+\ii} + \frac{1}{2\ell\cdot p_{12}+s_{12}}\Bigg\}~.
\]
Notice that both terms in this expression are UV finite. We can integrate the loop three-momentum for both contributions separately without the necessity of introducing a cut-off. The full analytic result is given by
\[
 {\cal A}_3^{(1)} = \frac{1}{32 \pi^2 s_{12}} \log^2 \left( \frac{\sqrt{s_{12} (s_{12}-4M^2)} + 2M^2-s_{12}}{2M^2} \right)~.
\]

The large mass expansion is straightforward and it is free of thresholds, i.e. the $\ii$ prescription can be dropped when $r=s_{12}/M^2\ll 1$. We need to consider both $G_D(q_2; q_3)$ and $G_D(q_3; q_2)$ in the context of the general propagator expansion. Since in both propagators the condition $\Gamma + \Delta = s_{12} < M \sqrt{s_{12}}$ holds we must identify $\Gamma=0$ and $\Delta =s_{12}$. Thus the asymptotic expansion of the propagators are given by
\[
	G_D(q_2; q_3) = \frac{1}{-2q_2\cdot p_{12} +s_{12}} = \sum_{n=0}^\infty \frac{\left( - s_{12} \right)^n}{\left( -2q_2\cdot p_{12} \right)^{n+1} }
\]
and
\[
	G_D(q_3; q_2) = \frac{1}{2q_3\cdot p_{12} +s_{12}} = \sum_{n=0}^\infty \frac{\left( - s_{12} \right)^n}{\left( 2q_3\cdot p_{12} \right)^{n+1} }~.
\]
Combining the two expanded propagators one obtains a single asymptotic expansion as
\[
	G_D(q_2; q_3) + G_D(q_3; q_2) = -\frac{2}{s_{12}} \sum_{n=1}^\infty  \left( \frac{s_{12}}{2\ell\cdot p_{12}} \right)^{2n}~,
\]
leading to the expanded amplitude
\[
{\cal A}_3^{(1)} ( s_{12} \ll M^2 )  =  - \int_\ell \frac{\td{\ell; M}}{(2\ell\cdot p_{12})(\ell\cdot p_1)} \sum_{n=1}
\left(\frac{s_{12}}{2\ell\cdot p_{12}} \right)^{2n}~.
\label{eq:trianglelarge}
\]
Integration leads to the following result for the large mass expansion:
\[
	{\cal A}_3^{(1)} ( s_{12} \ll M^2 )  
 	&= -\frac{1}{16\pi^2 } \frac{1}{2M^2} \left( 1+\frac{r}{12} + \frac{r^2}{90} \right) + {\cal O} (r^3)~. 
\]
For $M/\sqrt{s_{12}}=3$ the relative error of the result is $9\cdot 10^{-3}$ with only the first term of the expansion and reduces to $1\cdot 10^{-4}$ and $2\cdot 10^{-6}$ when including up the second and third term of the expansion, respectively.

In the small mass limit, $2M/\sqrt{s_{12}} \ll 1$, the general expansion of the dual propagators can be applied as well. The expansion parameters are
\[
	\Gamma_{32}=\Gamma_{23} \equiv \Gamma = s_{12} \left( 1 + \frac{M^2}{s_{12}} \right)~, \qquad 
	r_{32} = - r_{23} = - \frac{M}{\sqrt{s_{12}}} + \ii~, \qquad Q_2^2 = Q_3^2=s_{12}~.
\]
This leads to the expanded amplitude
\[
	{\cal A}_3^{(1)} (s_{12} \gg M^2 ) =  \, \int_\ell \frac{\td{\ell; M} s_{12}}{(2\ell\cdot p_{12})(2\ell\cdot p_1)} \sum_{n=0}^\infty \Bigg\{ 
 	\frac{M^{2n}}{\left( -2\ell\cdot p_{12} +\Gamma\right)^{n+1}}
	+ \frac{M^{2n}}{\left( 2\ell\cdot p_{12} +\Gamma\right)^{n+1}}\Bigg\}~. 
	\label{eq:trianglesmallreg} 
\]

Alternatively, both propagators in the sum may be combined as 
\[
	{\cal A}_3^{(1)} 
	=  \int_\ell \frac{\td{\ell; M}}{(2\ell\cdot p_{12})(2\ell\cdot p_1)} \frac{2 s_{12}^2}{(-\left( 2 \ell\cdot p_{12} \right)^2 + s_{12}^2 + \ii )}~,
\]
and expanded similarly to the general expansion so as to obtain the final expanded form at integrand level
\[
	{\cal A}_3^{(1)} (s_{12}\gg M^2) = \int_\ell \frac{\td{\ell; M}\, s_{12}^2 }{(2\ell\cdot p_{12})(\ell\cdot p_1)} 
	\sum_{n=0}^\infty \frac{\left(s_{12}^2 r_{23}^2 \, (2+r_{23}^2) \right)^{n}}{\left(-(2\ell\cdot p_{12})^2+\Gamma^2 \right)^{n+1}}~. 
	\label{eq:trianglesmall} 
\]
Also in this variation of the expansion it is possible to simplify the denominator in terms of the integration variable $x$ by making use of 
\[
- (2 \ell\cdot p_{12})^2 + s_{12}^2\, (1+r_{23}^2)^2 = s_{12}^2 \,  (x^2-r_{23}^2) (x^{-2}-r_{23}^2)~.
\]
Analytic integration up to $n=1$ gives the result
\[
	{\cal A}_3^{(1)} (s_{12}\gg M^2) &= \frac{1}{16\pi^2} \frac{1}{2(1+r_{23}^2)^2 s_{12}} \Bigg( \log^2 \left(-r_{23}^2 \right)  \\
 	&+ \frac{r_{23}^2 (r_{23}^2+2) \ln{-r_{23}^2}}{1+r_{23}^2} \left( \frac{2}{1-r_{23}^2} + \frac{\ln{-r_{23}^2}}{1+r_{23}^2} \right) + \mathcal{O} \left( r_{23}^4 \right) \Bigg)~. \nonumber
\]
Using the numerical values $\sqrt{s_{12}}/M=3$ the relative error of this result is $33\%$ ($7.5\%$) in the real (imaginary) part including only the first term of the expansion and reduces to $7.5\%$ ($0.04\%$) and $1.5\%$ ($0.26\%$) when including up the second and third term of the expansion, respectively. The relative errors obtained by integrating \eq{eq:trianglesmallreg} are slightly better but of the same order of magnitude with $26\%$ ($2.7\%$) with only the first term and $3.0\%$ ($1.3\%$) and $0.31\%$ ($0.27\%$) when including up the second and third term of the expansion, respectively. Even better results can be obtained by obtaining the parameters through expansion of the singularity position of the full propagator. In this case already at first order the relative error is at $3.9\%$ ($0.83\%$).

\section{Conclusions and Outlook}

In this work, we have presented the first results towards a general method for asymptotic expansions of Feynman amplitudes 
in the loop-tree duality formalism. The asymptotic expansion takes place at integrand-level in the Euclidean space of the 
loop three-momentum, where the hierarchies among internal and external scales are clear. 
The method is well-defined since convergence to the full integral is not only achieved in the final result but also at integrand-level, 
giving ample justification for applying these expansions. Additionally, the UV behaviour of the individual contributions to the asymptotic expansion 
does not increase when including more orders in the expansion. Renormalization is completed locally in four space-time dimensions 
with only the first terms of the expansion, and is not affected by increasing the precision through adding more orders. 
Both of these aspects are an improvement compared to the commonly used method of Expansion by Regions.

We have presented explicit results for the scalar two- and three-point functions at one loop in different kinematical limits. 
Specifically, we have achieved with a single expression a universal description of several asymptotic limits of the two-point function by 
conveniently selecting certain parameter of this expression.
More work is needed to make a wide range of applications possible at one loop and higher orders leading to additional challenges.
Recent developments in the realization of the LTD representation to all orders~\cite{Verdugo:2020kzh} could facilitate this task. 
Further results and physical applications are underway and will be published in forthcoming publications.

\acknowledgments

The project that gave rise to these results received the support of a fellowship from ``la Caixa`` Foundation (ID 100010434). The fellowship code is LCF/BQ/IN17/11620037. This work is supported by the Spanish Government  (Agencia Estatal de Investigaci\'on) and ERDF funds from European Commission (Grant No. FPA2017-84445-P), Generalitat Valenciana (Grant No. PROMETEO/2017/053), Consejo Superior de Investigaciones Cient\'{\i}ficas (Grant No. PIE-201750E021) and the COST Action CA16201 PARTICLEFACE.


\end{document}